\documentstyle[pra,aps,psfig,preprint,tighten]{revtex}
\begin{document}
\preprint{}
\draft
\title{Continuous Quantum Monitoring of Position  \\
of Nonlinear Oscillators} 
\author{Michael B. Mensky} 
\address{\em P. N. Lebedev Physical Institute, USSR Academy of Sciences, 
Moscow 117924, USSR}
\author{Roberto Onofrio} 
\address{\em INFN, Sezione di Roma, P.le Aldo Moro 2, Roma 00185, Italy} 
\author{Carlo Presilla}
\address{\em Consorzio INFM, Dipartimento di Fisica dell'Universit\`a 
di Perugia, Perugia 06100, Italy}
\date{Phys. Lett. A 161 (1991) 236-240}
\maketitle
\begin{abstract} 
Application of the path-integral approach to continuous measurements 
leads to effective Lagrangians or Hamiltonians in which the effect 
of the measurement is taken into account through an imaginary term. 
We apply these considerations to nonlinear 
oscillators with use of numerical computations to evaluate quantum 
limitations for monitoring position in such a class of systems.  
\end{abstract}
\pacs{ }

Several numerical experiments have shown that quantum-classical 
correspondence in chaotic dynamics seems to be anomalous [1]. 
In particular, it has been suggested that nonlinear systems showing chaotic 
behaviour in a classical regime may exhibit suppression of their
stochasticity when quantization is taken into account [2].
However this suppression should appear on a time scale so short that 
the classical chaotic behaviour could not be observed in real 
systems. To recover the Liouville 
dynamics use of quantum measurements process 
has been suggested [3]. In [4] an example of a chaotic system 
in a quantum regime under instantaneous measurements 
has been investigated. 

The quantum measurement processes for 
continuous monitoring of some observables of a dynamical system 
may be taken into account through path-integral formalism [5-7] and 
offers the possibility to investigate in a general way suppression of quantum 
behaviour. 
It is therefore interesting to apply this method to nonlinear systems, 
specifically those possessing chaotic properties in the classical 
regime, having as main goal the study of quantum suppression 
of chaotic behaviour in  such a class of systems. 
In this letter we deal with the development 
of a technique based upon numerical integration of an 
effective Schr\"odinger equation and 
applicable to continuous quantum measurements in nonlinear systems.
As an example a nonlinear oscillator will be investigated. 

The path-integral approach to continuous measurements is essentially based 
upon restriction of Feynman path integrals [5]. The integration is 
restricted to the set of paths compatible with the informations, namely the 
given output of the continuous measurement. 
If the system moves between the points of the space-time $(x^{\prime},0)$ 
and $(x^{\prime\prime},\tau)$ the result of integration 
$K_{[a]}(x^{\prime\prime},\tau;x^{\prime},0)$ depends on the measurement output 
$[a]=\{a(t)~|~0 \leq t \leq \tau \}$. This can be interpreted in two different ways. Firstly, it 
represents the probability amplitude for the measurement output 
$a(t)$ in the time interval $\tau$ 
given the positions of the system $x^{\prime}$ and $x^{\prime\prime}$ 
before and after the measurement. Secondly, it can be seen as a 
propagator for the system subject to the continuous measurement 
given the output of the measurement. 
The restriction of the path-integral on some 
set of paths compatible with the measurement output can be effectively done 
with the help of a weight functional $w_{[a]}[x]$ depending on the 
measurement output $a(t)$ and decaying outside the set of paths 
compatible with $a(t)$
\begin{equation}
K_{[a]}(x^{\prime\prime},\tau;x^{\prime},0)=
\int d[x] \exp \biggl\{ {i \over {\hbar}}\int_0^{\tau}L(x,\dot{x},t)dt \biggr\}
 w_{[a]}[x]
\end{equation}
 For example, if the coordinate $x$ is monitored with error $\Delta a$ 
and the result of the measurement is $a(t)$, the weight functional 
$w_{[a]}[x]$ selects those paths $x(t)$ in the corridor centered 
around $a(t)$ and having a width $\Delta a$ (see Fig. 1). 
A simple choice is a Gaussian weight functional
\begin{equation}
w_{[a]}[x]=\exp \biggl \{-{1 \over {\tau\Delta a^2}}\int_0^{\tau} 
(x(t)-a(t))^2 dt \biggr \}
\end{equation}
The restricted path integral is then 
\begin{equation}
K_{[a]}(x^{\prime\prime},\tau;x^{\prime},0)=\int d[x] \exp \biggl \{ 
{i \over {\hbar}}
\int_0^{\tau}L(x,\dot{x},t)dt-{1 \over {\tau \Delta a^2}}
\int_0^{\tau}(x-a)^2 dt \biggr \}
\end{equation}

It is important to observe that the resulting path integral may be considered 
as describing a free (i.e. not measured) system but with effective 
Lagrangian having an imaginary term due to the measurement [8] 
\begin{equation}
L_{eff}(x,\dot{x},t)=L(x,\dot{x},t)+{{i\hbar} \over {\tau \Delta a^2}}
(x-a(t))^2
\end{equation}
 This contribution produces a decrease of 
the density of the system in the configuration space far from $x(t)=a(t)$.
The decrease is linked with the restriction of the alternatives 
due to the measurement performed [9].
Finally the effective Lagrangian is time-dependent even 
if the original Lagrangian is not.

In order to estimate a probability distribution for the measurement 
output of a physical system we consider the convolution 
\begin{equation}
I_{[a]}=\langle\phi_2\vert K_{[a]}\vert\phi_1\rangle=
\int\int \phi_2^*(x^{\prime\prime}) K_{[a]}(x^{\prime\prime},\tau;
x^{\prime},0)\phi_1(x^{\prime}) dx^{\prime}dx^{\prime\prime}
\end{equation}
According to the first interpretation of $K_{[a]}$ the  
quantity $I_{[a]}$ is a probability amplitude for the measurement to give 
the output $a(t)$ under the condition that the system has been in the 
state $\phi_1$ before the measurement and in the state 
$\phi_2$ after a time $\tau$.
The probability distribution for the measurement output is then
\begin{equation}
P_{[a]}={{\vert I_{[a]}\vert}^2 \over {\int d[a]{\vert I_{[a]}\vert}}^2}
\end{equation}
Note that $P_{[a]}$ depends upon the instrumental uncertainty $\Delta a$.

The amplitude $I_{[a]}$ 
can be also written as a scalar product 
\begin{equation}
I_{[a]}=\langle\phi_2\vert\psi_{[a]}(\tau)\rangle
\end{equation}
where 
\begin{equation}
\psi_{[a]}(x^{\prime\prime},\tau)={\int} K_{[a]}(x^{\prime\prime},\tau;
x^{\prime},0)\phi_1(x^{\prime})dx^{\prime}
\end{equation}
According to the second interpretation of $K_{[a]}$ the 
wave function $\psi_{[a]}(x,t)$ represents the evolution at 
time $t$ of the state $\phi_1(x)$ under the action of a continuous 
measurement with output $a(t)$. This also means that $\psi_{[a]}(x,t)$ 
can be found as the solution of the time-dependent Schr\"odinger 
equation 
\begin{equation}
i\hbar{{\partial \psi_{[a]}(x,t)} \over {\partial t}}=H_{eff}\psi_{[a]}(x,t)
\end{equation}
with an effective Hamiltonian $H_{eff}$ corresponding to the 
Lagrangian $L_{eff}$ in (4) and with the choice 
$\psi_{[a]}(x,0)=\phi_1(x)$.

The time dependence of the effective Hamiltonian $H_{eff}$ makes 
analytical calculations difficult. In this case one may more simply 
evaluate $P_{[a]}$ through the Feynman propagator $K_{[a]}$. 
However, due to the quadratic nature of the measurement contribution 
to $L_{eff}$, analytical calculations are essentially restricted 
to the case in which $L(x,{\dot{x}},t)$ is a linear oscillator. 
In general a numerical approach 
must be followed and in this case the Schr\"odinger formalism is more 
suitable. The partial differential equation (9) is reduced to 
a simple finite difference recursive equation 
by choosing a proper lattice to simulate the continuous 
space-time [10]. No particular problems arise from the time dependence 
and the non-Hermitian nature of the differential operator $H_{eff}$ [11].

In order to understand the behaviour of a nonlinear system let us first 
consider a linear oscillator 
\begin{equation}
L={{m} \over {2}} {\dot{x}}^2-{{m\omega^2} \over {2}}x^2
\end{equation}
In this case analytical results have been obtained allowing a test of the  
numerical technique. The effective Lagrangian 
corresponds to a forced linear oscillator 
\begin{equation}
L_{eff}={m \over 2}{\dot{x}}^2-{{m\omega^2_{r}} \over {2}}x^2
-{{2i\hbar} \over {\tau\Delta a^2}}a(t)x+{{i\hbar} \over 
{\tau\Delta a^2}}a(t)^2
\end{equation}
with renormalized complex frequency 
\begin{equation}
\omega_{r}^2=\omega^2-{{2i\hbar} \over {m\tau\Delta a^2}}
\end{equation}
For any choice of the measurement output $a(t)$ the propagating 
kernel $K_{[a]}$ can be easily calculated [12]. Let us consider 
what happens when the quantum system is in the ground state of the 
unmeasured oscillator before and after the period $\tau$ of the 
continuous measurement
\begin{equation}
\phi_1(x)=\phi_2(x)={{ \biggl ({m\omega \over \pi\hbar} \biggr )^{1/4}}
\exp\biggl (-{{m\omega} \over {2\hbar}}x^2}\biggr )
\end{equation}
Due to the shape of $\phi_1$ and $\phi_2$ it is natural 
to choose the null boundary conditions $a(0)=a(\tau)=0$ for the 
measurement output $a(t)$. 
Any such function $a(t)$ can be written as a Fourier sine series.   
We consider only measurement outputs of the form
\begin{equation}
a(t)=\epsilon 
\sin \Omega_n t\ \ \ \ \ ; \ \ \ \ \ \ \Omega_n=n{{\pi} \over {\tau}}
\end{equation}
where $n$ is an integer number.
For a fixed $\Omega_n$ the probability distribution $P_{[a]}$ 
is reduced to a function of the amplitude $\epsilon$. 
$P(\epsilon)$ is a Gaussian function 
of width $\Delta a_{eff}$
\begin{equation}
P(\epsilon)={{1} \over {\sqrt{\pi}\Delta a_{eff}}}
\exp\biggl ( {-{{\epsilon^2} \over {\Delta a_{eff}^2}}}\biggr )
\end{equation}
where
$$ \Delta a_{eff}^{-2}=2\ \Re e\biggl \{ {{1} 
\over {2\Delta a^2}}\biggl [1-{{2i\hbar} \over
{m\tau\Delta a^2(\Omega_n^2-\omega^2_r)}}\biggr ]-$$
\begin{equation}
-{{4\hbar\Omega_n^2} \over 
{m \omega \tau^2 \Delta a^4(\Omega_n^2-\omega^2_r)^2}}
{\biggl [1-i{{\omega_r} \over {\omega}}\biggl (\cot (\omega_r\tau)+
{{(-1)^n} \over {\sin(\omega_r\tau)}}\biggr ) \biggr ]^{-1}\biggr \}}
\end{equation}

 The meaning of $\Delta a_{eff}$ 
is linked to the role of the measurement device during the evolution 
of the system under measurement. 
When the instrument error $\Delta a$ is large in comparison to the 
characteristic quantum scale of the system under measurement, 
$\sqrt{{\hbar} / {m\omega}}$ in this case,  the corridor shown in 
Fig. 1 contains the classical trajectory (with null boundary 
conditions). 
Thus the classical limit is  
\begin{equation}
\Delta a_c\equiv\lim_{\Delta a\rightarrow {\infty}} \Delta a_{eff}=\Delta a
\end{equation}
On the other hand when $\Delta a$ becomes small quantum noise arises. 
Also corridors far from the classical trajectory are probable. 
The quantum limit is written as 
\begin{equation}
\Delta a_q\equiv\lim_{\Delta a\rightarrow 0}
\Delta a_{eff}={\biggl [\biggl 
({m \over \hbar}\biggr )^{3 \over 2}\tau^{1 \over 2}\Omega_n^2
\Delta a+\biggl ({m\tau \over 2\hbar}\biggr )^2(\Omega_n^2-\omega^2)^2
\Delta a^2\biggr ]}^{-1/2}
\end{equation}
i.e. the effective error $\Delta a_{eff}$ diverges as 
$\Delta a^{-1/2}$. In an intermediate situation $\Delta a_{eff}$ 
interpolates between these two limits always maintaining values larger 
than the instrumental error. 
In Fig.~2 the behaviour of $\Delta a_{eff}$ versus $\Delta a$ is shown 
for three different values of the ratio $\Omega_n/\omega$.
As shown in (17) and (18), 
while the classical limit is the same for all the situations, 
different behaviours appear in the quantum regime. 
The minimum effective uncertainty 
is maximum at the resonance condition $\Omega_n=\omega$. 
In this case  it is always 
$\Delta a_{eff}\geq {\sqrt{{\hbar} / {m\omega}}}$. 
When $\Omega_n\neq\omega$ the minimum effective uncertainty 
estimated by the intersection between quantum and classical limits 
decreases as $(\Omega_n^2-\omega^2)^{-1}$.

The comparison between analytical and  
numerical results for $\Delta a_{eff}$ are shown in Fig.~3 
for two different choices of the measurement parameters. 
Also shown are classical and quantum behaviours, 
corresponding to the limits expressed in (17) and (18). 
The difference between numerical and analytical results is 
less than $0.1\%$ and it can be further reduced by choosing 
higher resolution space-time lattices.

The numerical accuracy estimated above allows us to 
perform meaningful computations
for a nonlinear oscillator represented by the Lagrangian
\begin{equation}
L={m \over 2}{\dot{x}}^2-{{m\omega^2} \over {2}}x^2-
{{\beta} \over {4}}x^4
\end{equation}
For comparison with the previous linear case we have chosen 
the initial and final states of the form (13) and a measurement 
output of the form (14). 
 In this case we expect both non-Gaussian behaviours for 
the wave function at $t=\tau$ and the propagator. 
This also means that the distribution $P(\epsilon)$ is not 
a Gaussian function. An  equivalent width for $P(\epsilon)$ 
may be introduced through the definition
\begin{equation}
\Delta a_{eff}\equiv {1 \over {\sqrt{\pi}P(0)}}\int_{-\infty}^{+\infty} 
P(\epsilon)d\epsilon
\end{equation}
The computed $\Delta a_{eff}$ versus $\Delta a$ 
are shown in Fig. 4 for two different values of 
the nonlinearity coefficient $\beta$. The comparison with the 
linear situation having the same parameters is also shown. It appears 
that the effect of the nonlinear term is in the direction to 
enlarge the region in which the classical approximation is meaningful. 
The quartic term concentrates the final wave function near the 
result of the measurement and this implies that the effect of 
quantum noise, toward a spreading of the most probable paths, is reduced.

In all the previous considerations we have chosen to deal with 
$\Delta a$ which is time independent. A more general class of 
continuous measurements is obtained by considering time 
dependence for $\Delta a$. In particular to recover chaotic dynamics 
through quantum measurements a particular conditions on the kind 
of measurements must be satisfied, namely the measurement process 
has to be a quantum nondemolition process for the observable 
under monitoring [4]. Quantum nondemolition strategies for linear
oscillators have been already analyzed in the framework of the 
path integral with continuous measurements [8].
The application of the technique described 
here for quantum nondemolition measurement processes on 
nonlinear systems will be the subject of future investigations. 
\vskip 1.0truecm
Acknowledgements: One of us (M.B.M.) is indebted to Prof. G. Immirzi 
and Prof. F. Marchesoni for the kind hospitality in the University 
of Perugia where a part of this work has been completed.
\vskip 1.5truecm

{\bf References}

[1] G. Casati, B. V. Chirikov, F. M. Izrailev and J. Ford, in 
{\sl Stochastic Behavior in Classical and Quantum Hamiltonian 
Systems}, edited by G. Casati and J. Ford
 (Springer-Verlag, New York, 1979), p. 334.

[2] B. V. Chirikov, F. M. Izrailev, D. L. Shepelyansky, 
{\sl Sov. Sci. Rev. Sect.} {\bf C2} (1981) 209.

[3] D. L. Shepelyansky, {\sl Physica} {\bf 8D} (1983) 208.

[4] S. Adachi, M. Toda and K. Ikeda, {\sl Journ. Phys. A} {\bf 22} (1989) 3291.

[5] M. B. Mensky, {\sl Phys. Rev. D} {\bf 20} (1979) 384; 
{\sl Sov. Phys. JETP} {\bf 50} (1979) 667.

[6] F. Ya Khalili, {\sl Vestnik Mosk. Univers.}, ser. 3 {\bf 22} (1981) 37; 
A. Barchielli, L. Lanz and G. M. Prosperi, {\sl Nuovo Cimento B} {\bf 72} 
(1982) 79;
C. M. Caves, {\sl Phys. Rev. D} {\bf 33} (1986) 1643; {\sl Phys. Rev. D} 
{\bf 35} (1987) 1815.

[7] M. B. Mensky, {\sl Phys. Lett. A} {\bf 155} (1991) 229.

[8] G. A. Golubtsova and M. B. Mensky, {\sl Int. Journ. Mod. Phys. A} {\bf 4} 
(1989) 2733.

[9] For a detailed discussion of this effect and the 
related irreversibility introduced during the measurement 
see M. B. Mensky, {\sl Theor. Math. Phys.} {\bf 75} (1988) 357.

[10] W. H. Press, B. P. Flannery, S. A. Teukolsky and W. T. Vetterling, 
{\sl Numerical Recipes: the Art of Scientific Computing} (Cambridge 
Univ. Press, Cambridge, 1986).

[11] C. Presilla, Doctoral Thesis, University of Rome, 1990 (unpublished); 
     C. Presilla, G. Jona-Lasinio and F. Capasso, {\sl Phys. Rev. B} {\bf 43} 
(1991) 5200.

[12] R. P. Feynman and H. R. Hibbs, {\sl Quantum Mechanics and Path Integrals} 
(McGraw-Hill, New York, 1965).

\begin{figure}   
\centerline{\hbox{\psfig{figure=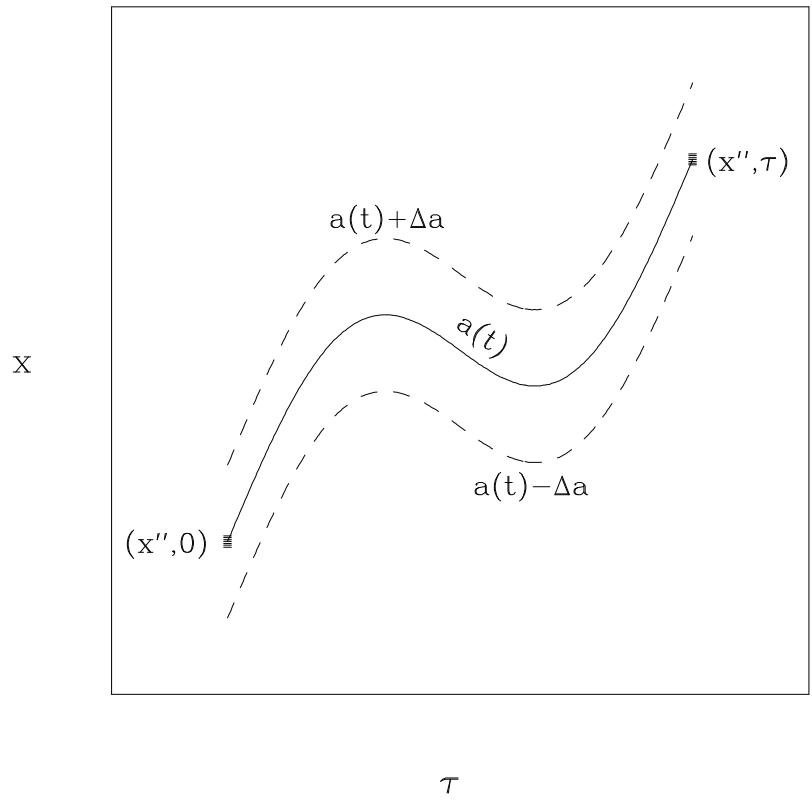,width=17cm,angle=90}}}
\caption{
A continuous measurement of position leads to a corridor 
centered at the measurement result $a(t)$ where the most probable paths lie.
}\end{figure}\noindent

\begin{figure}   
\centerline{\hbox{\psfig{figure=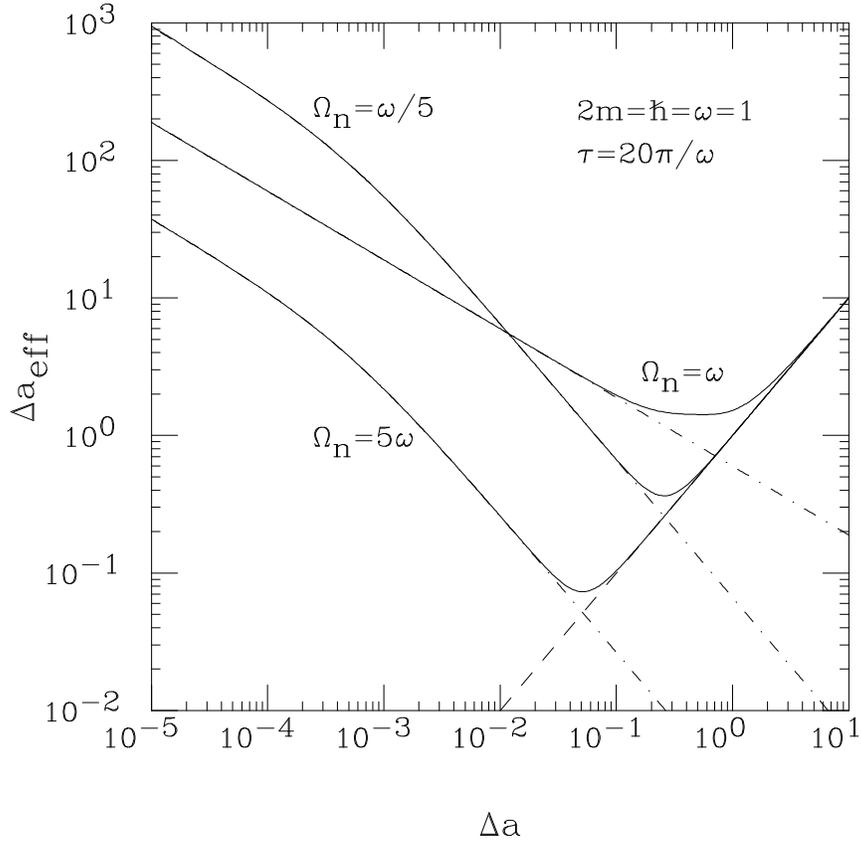,width=17cm,angle=90}}}
\caption{
Effective uncertainty versus instrumental uncertainty for three  
different values of the ratio $\Omega_n/\omega$ in the case of a linear 
oscillator. The dashed 
line is the classical limit $\Delta a_c$, the dot-dashed lines 
are the quantum limits $\Delta a_q$. The value of the other 
parameters is also indicated.
}\end{figure}\noindent

\begin{figure}   
\centerline{\hbox{\psfig{figure=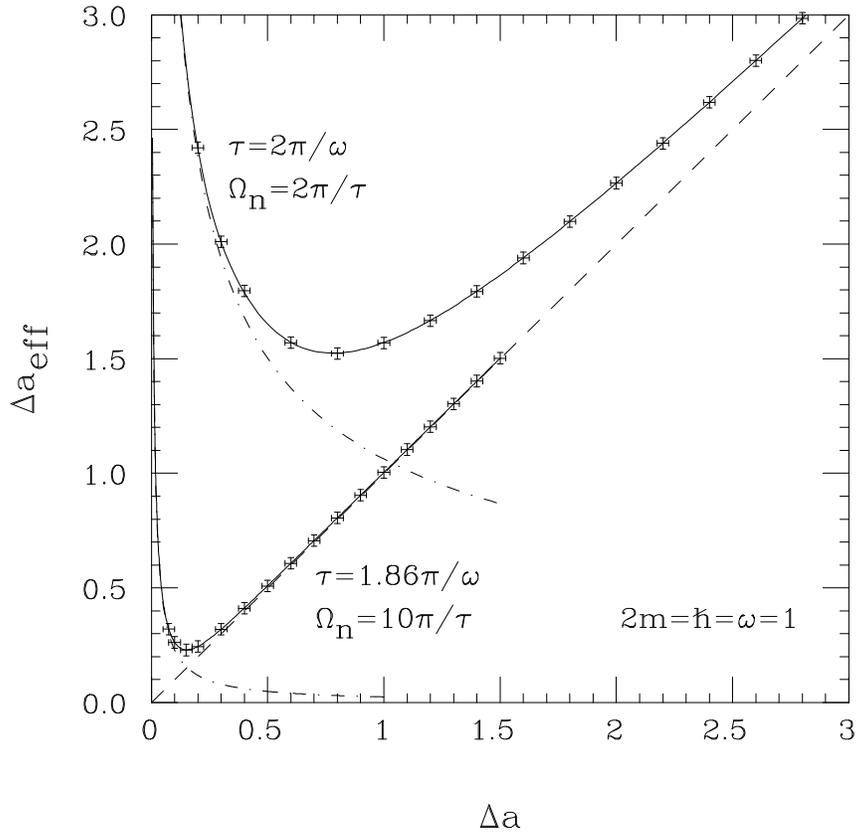,width=17cm,angle=90}}}
\caption{
Numerical (dots) and analytical (solid line) 
effective uncertainty versus 
instrumental uncertainty for a linear oscillator. 
Asymptotic classical (dashed) and quantum (dot-dashed) behaviours are 
also shown.
}\end{figure}\noindent

\begin{figure}   
\centerline{\hbox{\psfig{figure=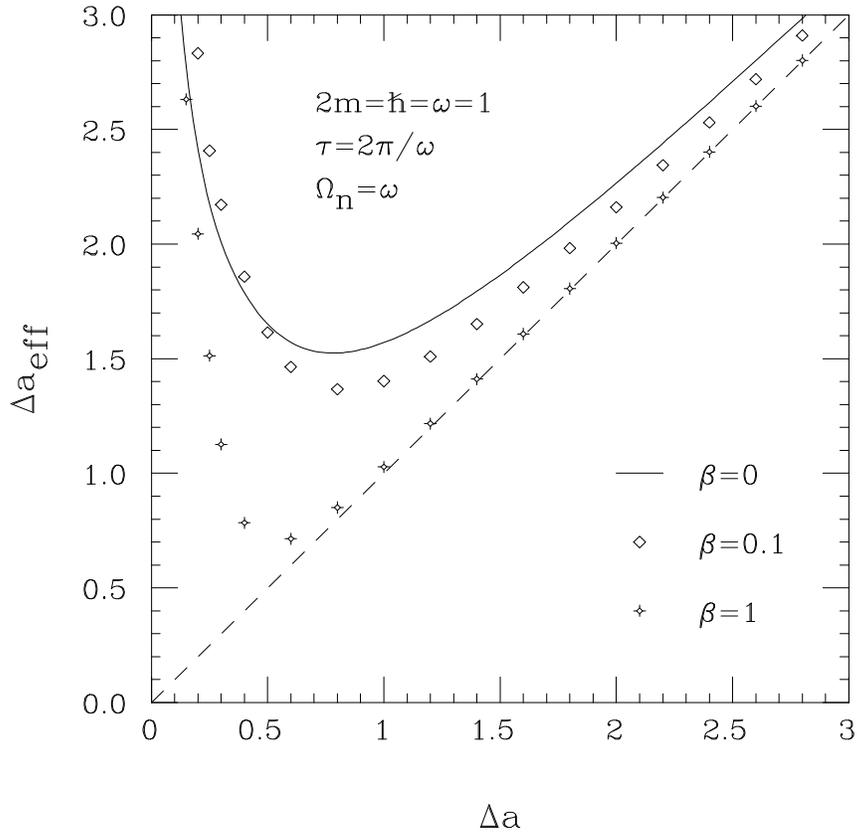,width=17cm,angle=90}}}
\caption{
Numerical (dots) effective uncertainty versus instrumental 
uncertainty for the nonlinear oscillator at two different values 
of $\beta$. The solid line is the behaviour of the 
corresponding linear oscillator obtained for $\beta=0$.
The dashed line represents the classical limit.  
}\end{figure}\noindent

\end{document}